\documentclass[twocolumn,aps,prc,showpacs,floatfix]{revtex4}
\usepackage{graphicx}
\usepackage{dcolumn}
\usepackage{bm}
\usepackage{mhchem}

\begin{document}

\title{Effects of symmetry energy in $^{132}\rm {Sn}+^{124}\rm {Sn}$ reaction at 300 MeV/nucleon}

\author{Shan-Jing Cheng$^{1,2}$}
\author{Gao-Chan Yong$^{2}$}\email{yonggaochan@impcas.ac.cn}
\author {De-Hua Wen$^{1}$}
\affiliation{%
$^1${School of Physics and Optoelectronic Technology, South China University of Technology, Guangzhou 510641, P.R. China}\\ $^2${Institute of Modern Physics,
Chinese Academy of Sciences, Lanzhou 730000, China}
}%

\date{\today}

\begin{abstract}
Based on the recently updated Isospin-dependent Boltzmann-Uehling-Uhlenbeck (IBUU) transport model, we studied the effects of symmetry energy on the neutron to proton n/p ratio and the $\pi^{-}/\pi^{+}$ ratio in central $^{132}\rm {Sn}+^{124}\rm {Sn}$ reaction at 300 MeV/nucleon. It is found that the n/p ratio and the $\pi^{-}/\pi^{+}$ ratio in central $^{132}\rm {Sn}+^{124}\rm {Sn}$ reaction at
300 MeV/nucleon mainly probe the symmetry energy in the density region 1-1.5 times saturation density. However, the $\pi^{-}/\pi^{+}$ ratio may be able to probe the density-dependent symmetry energy above 1.5 times saturation density by making some kinematic restrictions such as the azimuthal angle and kinetic energy cuts of emitting pions.

\end{abstract}


\maketitle

\section{Introduction}

The equation of state (EoS) of
nuclear matter at density $\rho$ and isospin asymmetry
$\delta$ ($\delta=(\rho_n-\rho_p)/(\rho_n+\rho_p)$) can be
expressed as \cite{li08,bar05}
\begin{equation}
E(\rho ,\delta )=E(\rho ,0)+E_{\text{sym}}(\rho )\delta ^{2}+\mathcal{O}%
(\delta ^{4}),
\end{equation}%
where $E_{\text{sym}}(\rho)$ is nuclear symmetry energy.
Nowadays the EoS
of isospin symmetric nuclear matter $E(\rho, 0)$ is relatively well
determined \cite{pawl2002} but the EoS of isospin
asymmetric nuclear matter, especially the high-density behavior
of the nuclear symmetry energy is still very uncertain \cite{Guo14},
e.g., there are plenty of analyses showing conflicting
results on pion production \cite{wolter06,xie13,xiao09,prassa07,feng10,hong2014,Reisdorf07,cozma16} when compared to FOPI data \cite{Reisdorf07} at GSI.
Constraints on the high-density behavior of the symmetry energy from ground-based measurements can be highly relevant to neutron stars \cite{Lat01}, such as their stellar radii and moments of inertia, crustal vibration frequencies \cite{Lat04,Vil04}, and neutron star cooling rates \cite{Lat04,Ste05}.

It is known to all that the so-called symmetry-energy-sensitive hadronic observables of probing the symmetry energy at supra-saturation densities in heavy-ion collisions inevitably suffer from affections of the symmetry energy at low densities as well as the final-state interactions, it is thus necessary to study the probed density region of some
symmetry-energy-sensitive observable more specifically \cite{liu15}.
Experimentally, related measurements of pions, nucleon, triton and $^{3}$He yield ratios
in $^{132}\rm {Sn}+^{124}\rm {Sn}$ reactions at
300 MeV/nucleon, are being carried out at RIBF-RIKEN in Japan \cite{sep} using the
SAMURAI-Time -Project-Chamber \cite{shan15}. And the main goal of the experiments is the
probe of the symmetry energy around twice saturation density.

In order to check whether the above measurements
can probe the symmetry energy around twice saturation density,
we in this work study the probed density region of the symmetry energy of some
symmetry-energy-sensitive observables in $^{132}\rm {Sn}+^{124}\rm {Sn}$ reaction at 300 MeV/nucleon. Our studies are based on the newly updated Isospin-dependent
Boltzmann-Uehling-Uhlenbeck (IBUU) transport model, which includes
physical considerations of nucleon short-range correlations, in-medium inelastic baryon-baryon cross section and momentum-dependent pion potential \cite{yong20151,yong20152,yong20153}.
Our studies show that neutron to proton ratio and $\pi^{-}/\pi^{+}$ ratio
in central $^{132}\rm {Sn}+^{124}\rm {Sn}$ reaction at
300 MeV/nucleon mainly probe the symmetry energy around saturation density, i.e., probe the symmetry energy in the density region 1-1.5 times saturation density. But the $\pi^{-}/\pi^{+}$ ratio may be able to probe the density-dependent symmetry energy at maximal density reached in heavy-ion collisions by making some kinematic restrictions of emitting $\pi$'s.

\section{A brief description to the
IBUU transport model}

The present used Isospin-dependent
Boltzmann-Uehling-Uhlenbeck (IBUU) transport model \cite{yong20151,yong20152} has its origin from the IBUU04 model \cite{lyz05}. In this model,
the proton and neutron
momentum distributions with high-momentum tail reaching about twice the Fermi momentum are used \cite{yong20151}.
The isospin- and momentum-dependent baryon mean-field
potential is expressed as
\begin{eqnarray}
U_B(\rho,\delta,\vec{p},\tau)&=&A_u(x)\frac{\rho_{\tau'}}{\rho_0}+A_l(x)\frac{\rho_{\tau}}{\rho_0}\nonumber\\
& &+B(\frac{\rho}{\rho_0})^{\sigma}(1-x\delta^2)-8x\tau\frac{B}{\sigma+1}\frac{\rho^{\sigma-1}}{\rho_0^\sigma}\delta\rho_{\tau'}\nonumber\\
& &+\frac{2C_{\tau,\tau}}{\rho_0}\int
d^3\,\vec{p^{'}}\frac{f_\tau(\vec{r},\vec{p^{'}})}{1+(\vec{p}-\vec{p^{'}})^2/\Lambda^2}\nonumber\\
& &+\frac{2C_{\tau,\tau'}}{\rho_0}\int
d^3\,\vec{p^{'}}\frac{f_{\tau'}(\vec{r},\vec{p^{'}})}{1+(\vec{p}-\vec{p^{'}})^2/\Lambda^2},
\label{buupotential}
\end{eqnarray}
where $\rho_0$ denotes saturation density, $\tau, \tau'=1/2(-1/2)$ for neutron potential $U_{n}$ (proton potential $U_{p}$),
$\delta=(\rho_n-\rho_p)/(\rho_n+\rho_p)$ is the isospin asymmetry,
and $\rho_n$, $\rho_p$ denote neutron and proton densities,
respectively. The specific values of the parameters used in the above baryon potential can be found in Ref. \cite{yong20152}. For baryon resonance $\Delta$ potential, the forms of
\begin{eqnarray}
U^{\Delta^-}_{B}&=& U_{n},\\
U^{\Delta^0}_{B}&=& \frac{2}{3}U_{n}+\frac{1}{3}U_{p},\\
U^{\Delta^+}_{B}&=& \frac{1}{3}U_{n}+\frac{2}{3}U_{p},\\
U^{\Delta^{++}}_{B}&=& U_{p}
\end{eqnarray}
are used. In these baryon potentials,
different symmetry energy's stiffness parameter $x$
can be used in the above baryon potentials to mimic
different forms of the symmetry energy.

The isospin-dependent baryon-baryon ($BB$) scattering cross section in medium $\sigma
_{BB}^{medium}$ is reduced compared with their free-space value
$\sigma _{BB}^{free}$ by a factor of
\begin{eqnarray}
R^{BB}_{medium}(\rho,\delta,\vec{p})&\equiv& \sigma
_{BB_{elastic, inelastic}}^{medium}/\sigma
_{BB_{elastic, inelastic}}^{free}\nonumber\\
&=&(\mu _{BB}^{\ast }/\mu _{BB})^{2},
\end{eqnarray}
where $\mu _{BB}$ and $\mu _{BB}^{\ast }$ are the reduced masses
of the colliding baryon-pair in free space and medium,
respectively. And the effective mass of baryon in isospin asymmetric nuclear matter
is expressed as
\begin{equation}
\frac{m_{B}^{\ast }}{m_{B}}=\left\{ 1+\frac{m_{B}}{p}\frac{%
dU_{B}}{dp}\right\}.
\end{equation}
In this model, the isospin-, density- and momentum-dependent pion potential
is also included \cite{yong20153}. More details about this model can be found in Ref. \cite{yong20151,yong20152,yong20153}.

%\section{Results and Discussions}

\begin{figure}[t]
\centering
\includegraphics [width=0.55\textwidth]{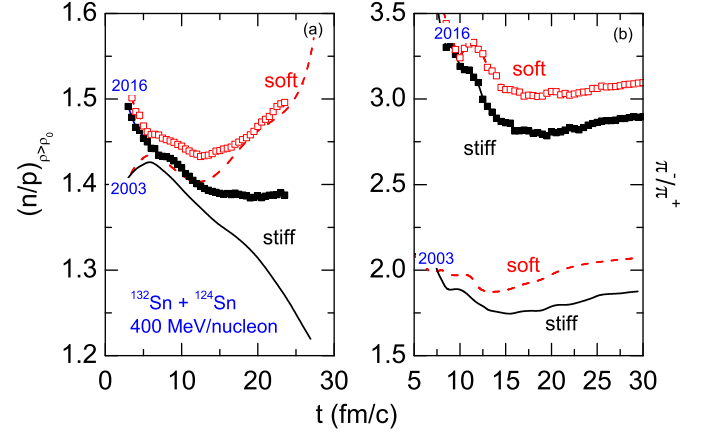}
\caption{(Color online) \label{compar} Comparisons of average neutron/proton ratio with densities higher than the normal nuclear matter density
as a function of time (Left panel) and $\pi ^{-}/\pi ^{+}$ ratio as a
function of time (Right panel) calculated with the IBUU model used in 2003 in Ref.~\cite{li2003} (shown with lines) and that with the IBUU model used in 2016 in Ref.~\cite{yong20152} (shown with symbols).
The solid (dashed) lines or solid (hollow) symbols are
the results using almost the same stiff (soft) symmetry energy.}
\end{figure}
Sensitivities of neutron to proton ratio and $\pi^-/\pi^+$ ratio to nuclear symmetry energy have in fact been examined by \emph{Li} in central $^{124}$Sn+$^{124}$Sn reactions at 400 MeV/nucleon in 2003 with an Isospin-dependent
Boltzmann-Uehling-Uhlenbeck (IBUU) transport model \cite{li2003}. Compared with the IBUU model used in 2003 in Ref.~\cite{li2003}, the newly updated IBUU model here considered the in-medium effects of nucleon-nucleon or baryon-baryon elastic and inelastic scatterings and the propagation of pion meson in medium. Moreover, we considered the effects of nucleon short-range correlations on nuclear initialization and kinetic symmetry energy calculation \cite{yong20152}. Of course, the momentum-dependence of single particle potential is also included \cite{yong20151,yong20152,yong20153}.

Shown in Fig.~\ref{compar} are comparisons of average neutron/proton ratio above normal nuclear matter density and $\pi ^{-}/\pi ^{+}$ ratio as a function of time in the same $^{124}$Sn+$^{124}$Sn reaction at 400 MeV/nucleon with different IBUU transport models but almost the same stiff or soft density-dependent symmetry energies.
In the left panel we compare the average $(n/p)_{\rho > \rho_0}$ of the whole space where the local densities are higher than the normal density $\rho_0$. With both IBUU models, effects of the symmetry energy on the $(n/p)_{\rho > \rho_0}$ is clearly seen when a sufficiently high compression has been reached. The value of $(n/p)_{\rho > \rho_0}$ with the IBUU of 2016 version is overall higher than that with the IBUU of 2003 version. The difference of predictions is more clearly seen in the right panel of Fig.~\ref{compar}, the $\pi ^{-}/\pi ^{+}$ ratio as a function of time. While keeping almost the same sensitivity of the $\pi ^{-}/\pi ^{+}$ ratio to the density-dependent symmetry energy, the value of $\pi ^{-}/\pi ^{+}$ ratio calculated with the IBUU used in 2003 is much lower than that calculated with the updated IBUU model in 2016.

The reasons of difference may from several aspects.
The used momentum dependence of the single particle potential in the IBUU2016 decreases the free neutron to proton ratio, which is a direct reflection of the decreasing symmetry potential with momentum \cite{lim2003}. The small value of free neutron to proton ratio causes high neutron to proton ratio of dense matter, which corresponding a high value of $\pi ^{-}/\pi ^{+}$ ratio \cite{gaoy2011}.
As discussed in Ref.~\cite{guom2014}, the in-medium cross section of pp (proton-proton) colliding pair is reduced more than that of nn (neutron-neutron) colliding pair. The effects of in-medium nucleon-nucleon scattering cross section on the $\pi ^{+}$ production is thus larger than that of $\pi ^{-}$. Therefore the in-medium effect increases the value of $\pi ^{-}/\pi ^{+}$ ratio \cite{guom2014}. While in fact in the IBUU2003, the free nucleon-nucleon cross section was used.
The asymmetry of reaction $^{132}\rm {Sn}+^{124}\rm {Sn}$ is roughly the same as that of Au+Au reaction system, and also the value of $\pi ^{-}/\pi ^{+}$ ratio decreases as the incident beam energy \cite{npa2010,mzhang2009}, the value of $\pi ^{-}/\pi ^{+}$ ratio in $^{132}\rm {Sn}+^{124}\rm {Sn}$ at 300 MeV/nucleon should be not smaller than that in Au+Au reaction at 400 MeV/nucleon. Since we give a reasonable prediction of $\pi ^{-}/\pi ^{+}$ ratio ($\pi ^{-}/\pi ^{+}$ $\simeq$ 3) in Au+Au reaction at 400 MeV/nucleon \cite{yong20152}, the present prediction of the value of $\pi ^{-}/\pi ^{+}$ ratio in $^{132}\rm {Sn}+^{124}\rm {Sn}$ at 300 MeV/nucleon should be also roughly correct.
And since the isospin-sensitive observables are sensitive to the detailed
tactics of transport model code, besides the differences of main inputs in
transport models, some specific details of algorithm may also affect the results.

\section{Results and Discussions}

\begin{figure}[t]
\centering
\includegraphics [width=0.5\textwidth]{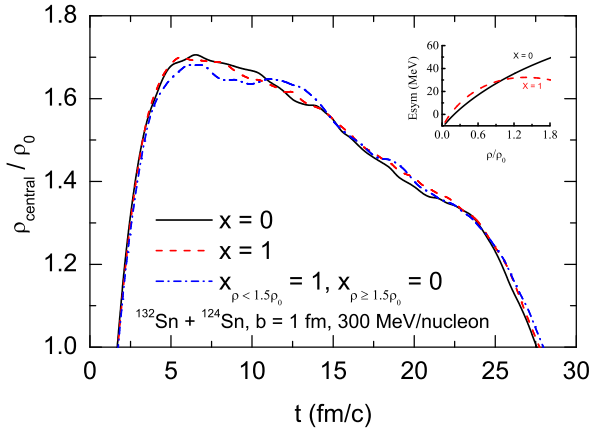}
\caption{(Color online) \label{dens} Maximal baryon density reached in central $^{132}\rm {Sn}+^{124}\rm {Sn}$ reaction at 300 MeV/nucleon with stiff (x = 0) or soft (x = 1) symmetry energies. The corresponding symmetry
energy functionals with x = 0 (stiff) and 1 (soft) are depicted in the inset. To check whether or not the symmetry energy at densities above 1.5$\rho_{0}$ affects symmetry-energy-sensitive observables, we also calculated the case x$_{\rho < 1.5\rho{_0}}$ = 1, x$_{\rho\geq1.5\rho{_0}}$ = 0 for comparison.}
\end{figure}
To probe the symmetry energy at supra-saturation densities, it is necessary to first show the maximal density reached in heavy-ion collisions. Shown in Fig.~\ref{dens} is evolution of the central baryon density in $^{132}\rm {Sn}+^{124}\rm {Sn}$ reaction at 300 MeV/nucleon with different symmetry energy settings. The corresponding symmetry
energy functionals with x = 0 (stiff) and 1 (soft) are depicted in the inset of Fig.~\ref{dens}. It is seen that in $^{132}\rm {Sn}+^{124}\rm {Sn}$ reaction at 300 MeV/nucleon, the maximal baryon density reached is about 1.7 times saturation density. Therefore, the central reaction of  $^{132}\rm {Sn}+^{124}\rm {Sn}$ at 300 MeV/nucleon may be not able to probe the high-density behavior of the symmetry energy around twice saturation density. To check whether or not the symmetry energy at densities above 1.5$\rho_{0}$ affects symmetry-energy-sensitive observables, we also calculated the case x$_{\rho < 1.5\rho{_0}}$ = 1, x$_{\rho\geq1.5\rho{_0}}$ = 0 for comparison.
The setting of changing x from 1 to 0 at the density point 1.5$\rho_{0}$
is just a mathematical trick to show whether or not the symmetry energy at densities above 1.5$\rho_{0}$ affects values of the n/p ratio or the $\pi ^{-}/\pi ^{+}$ ratio.

\begin{figure}[t]
\begin{center}
\includegraphics[width=0.5\textwidth]{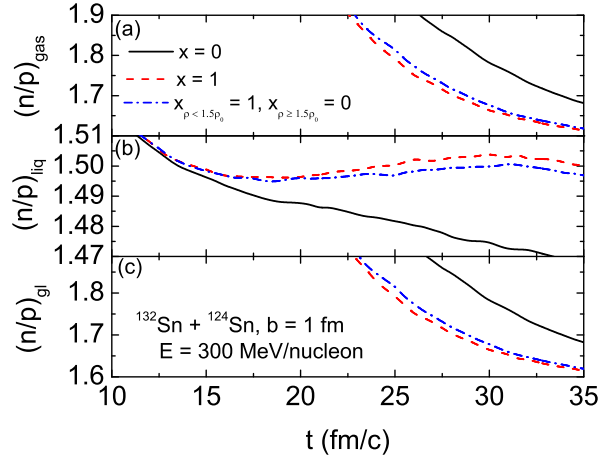}
\end{center}
\caption {(Color online) \label{frac} Evolution of neutron to proton ratio n/p in isospin fractionation. The n/p ratio of in gas phase is shown in panel (a). Panel (b) shows the n/p ratio of in liquid phase and panel (c) denotes the strength of isospin fractionation. We identify free nucleon by its local density $\rho \leq 1/8\rho_{0}$.}
\end{figure}
Nuclear matter formed in heavy-ion collisions is demonstrated to exhibit a phase transition
between a Fermi liquid and a nucleonic gas, and the gas phase is more enriched in neutrons than the liquid phase represented by bound nuclei \cite{xu2000}. To demonstrate the isospin fractionation in the central $^{132}\rm {Sn}+^{124}\rm {Sn}$ reaction at 300 MeV/nucleon, we analyzed the neutron to proton ratio n/p in isospin fractionation.
Fig.~\ref{frac} shows evolution of neutron to proton ratio n/p in isospin fractionation. We identify free nucleon by its local density $\rho \leq 1/8\rho_{0}$. From panel (a), we see that the value of n/p ratio of gas phase is larger than the value of the reaction system 1.56 and from panel (b) we see that the value of n/p ratio of liquid phase is smaller than the value of the reaction system 1.56.
The value of the n/p ratio of gas phase is thus larger than that of the liquid phase. From Fig.~\ref{frac}, one can clearly see that the value of the n/p ratio in isospin fractionation is evidently affected by the symmetry energy. For the stiffer symmetry energy x = 0, the value of the n/p ratio of gas phase is larger than that with the softer symmetry energy x = 1. The stiffer symmetry energy repels neutrons more strongly, thus more neutrons become free in the $^{132}\rm {Sn}+^{124}\rm {Sn}$ reaction. The larger the value of the n/p ratio of gas phase, the smaller the value of the n/p ratio of liquid phase.

From Fig.~\ref{frac}, one can also see that the effects of the symmetry energy above 1.5 times saturation density can be neglected in the $^{132}\rm {Sn}+^{124}\rm {Sn}$ reaction at 300 MeV/nucleon. Although the maximal density of compression matter can reach about 1.7 times saturation density as shown in Fig.~\ref{dens},
baryons have large probability lying in the density region below 1.5 times saturation density in the whole reaction process \cite{liu15}. This is the reason why the n/p ratio probes the symmetry energy below 1.5 times saturation density. Note here that the n/p ratio probes the symmetry energy of high density. We can see this point from pane (a), the value of the n/p ratio of gas phase with the stiff symmetry energy (x = 0) is larger than that with the soft symmetry energy (x = 1). Because there is a crossover of different density-dependent symmetry energies at the saturation density point as shown in the inset of Fig.~\ref{dens}, the value of n/p ratio of gas phase with the stiff symmetry energy (x = 0) should be smaller than that with the soft symmetry energy (x = 1) if isospin fractionation in $^{132}\rm {Sn}+^{124}\rm {Sn}$ reaction at 300 MeV/nucleon is mainly affected by the low-density behavior of the symmetry energy.

\begin{figure}[t]
\begin{center}
\includegraphics[width=0.5\textwidth]{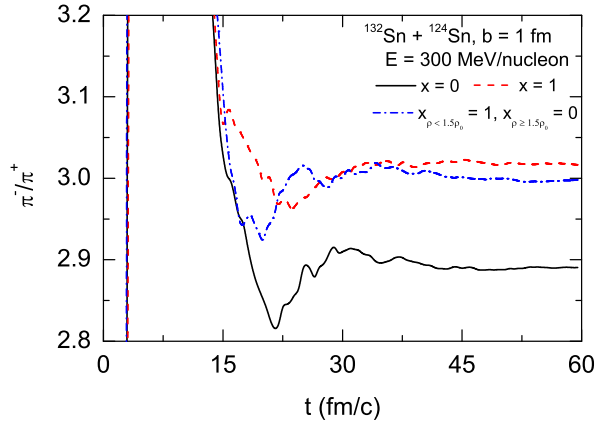}
\end{center}
\caption {(Color online) \label{rpion} Evolution of the $\pi ^{-}/\pi ^{+}$ ratio in the central $^{132}\rm {Sn}+^{124}\rm {Sn}$ reaction at 300 MeV/nucleon with different symmetry energy settings.}
\end{figure}
According to $\Delta $ resonance model
\cite{stock}, $\pi ^{-}/\pi ^{+}$ ratio is
approximately equal to $(5N^{2}+NZ)/(5Z^{2}+NZ)\approx (N/Z)^{2}$ in central
heavy-ion reactions with $N$ and $Z$ being the total neutron and proton
numbers in the participant region \cite{stock}. Since the neutron to proton ratio in dense matter formed in heavy-ion collisions is sensitive to the symmetry energy \cite{lyz05}, the $\pi^-/\pi^+$ ratio is a potential sensitive probe of the symmetry energy. With this in mind,
both the $\pi^-/\pi^+$ single \cite{lyz05} and double ratios \cite{ylcz06} were studied to show sensitivities to the high-density behavior of the
symmetry energy. But specific density region that the $\pi^-/\pi^+$ ratio probed in heavy-ion collisions was seldom studied until very recently \cite{liu15}.

Fig.~\ref{rpion} shows evolution of the $\pi ^{-}/\pi ^{+}$ ratio in central $^{132}\rm {Sn}+^{124}\rm {Sn}$ reaction at 300 MeV/nucleon with different symmetry energies. The $\pi ^{-}/\pi ^{+}$ ratio reads as \cite{li2003}
\begin{equation}
\pi^-/\pi^+\equiv \frac{\pi^-+\Delta^-+\frac{1}{3}\Delta^0}
{\pi^++\Delta^{++}+\frac{1}{3}\Delta^+}.
\end{equation}
From Fig.~\ref{rpion}, one sees that the soft symmetry energy (x = 1) corresponds to a larger value of the $\pi ^{-}/\pi ^{+}$ ratio while the stiff symmetry energy (x = 0) corresponds to a smaller value. This is consistent with the studies in Refs. \cite{lyz05,ylcz06,ono16}. Again, one sees that the effects of the symmetry energy above 1.5 times saturation density can be neglected. This is understandable from the n/p ratio of liquid phase shown in panel (b) of Fig.~\ref{frac}. The stiff symmetry energy (x = 0) corresponds to a low value of the n/p ratio of liquid phase, thus less (N/Z)$^{2}$ value in the participant region. A small (N/Z)$^{2}$ value in the participant region corresponds to a small value of the $\pi ^{-}/\pi ^{+}$ ratio. As similar discussions in Fig.~\ref{frac}, here the
$\pi ^{-}/\pi ^{+}$ ratio in central $^{132}\rm {Sn}+^{124}\rm {Sn}$ reaction at 300 MeV/nucleon still probes the high-density behavior of the symmetry energy. The result that the
$\pi ^{-}/\pi ^{+}$ ratio probes the symmetry energy in the density region 1-1.5 times saturation density in central $^{132}\rm {Sn}+^{124}\rm {Sn}$ reaction at 300 MeV/nucleon is consistent with the recent study in Ref. \cite{liu15}, while the latter study is based on a transport model with the momentum-independent baryon potential and the free baryon-baryon inelastic cross section and without considering nucleon short-range correlations \cite{yong20152}.

\begin{figure}[t]
\begin{center}
\includegraphics[width=0.5\textwidth]{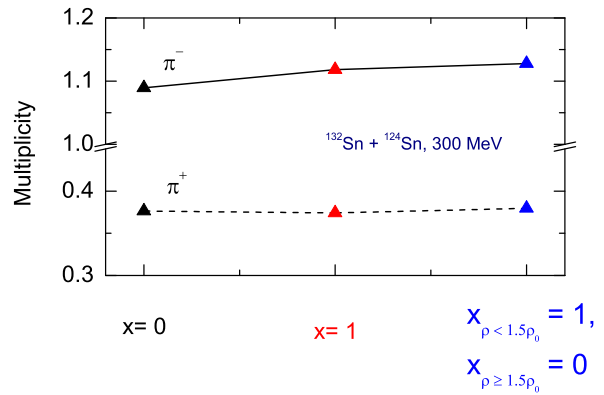}
\end{center}
\caption {(Color online) \label{totpion} Average multiplicity of $\pi ^{-}$, $\pi ^{+}$ as
a function of x parameter in the reaction $^{132}$Sn + $^{124}$Sn at a
beam energy of 300 MeV/nucleon with an impact parameter of 1 fm.}
\end{figure}
Fig.~\ref{totpion} shows average multiplicity of $\pi ^{-}$, $\pi ^{+}$ as
a function of x parameter in the reaction $^{132}$Sn + $^{124}$Sn at a
beam energy of 300 MeV/nucleon with an impact parameter of 1 fm. It is seen that the yield of $\pi ^{-}$ is about 3 times that of $\pi ^{+}$ in central $^{132}$Sn + $^{124}$Sn reaction at a
beam energy of 300 MeV/nucleon. And $\pi ^{-}$ production is sensitive to the symmetry energy while $\pi ^{+}$ production is not sensitive to the symmetry energy. The number of produced $\pi ^{-}$ ($\pi ^{+}$) is about 1.1 (0.38) per event in the central $^{132}$Sn + $^{124}$Sn reaction at 300 MeV/nucleon.

\begin{figure}[t]
\begin{center}
\includegraphics[width=0.5\textwidth]{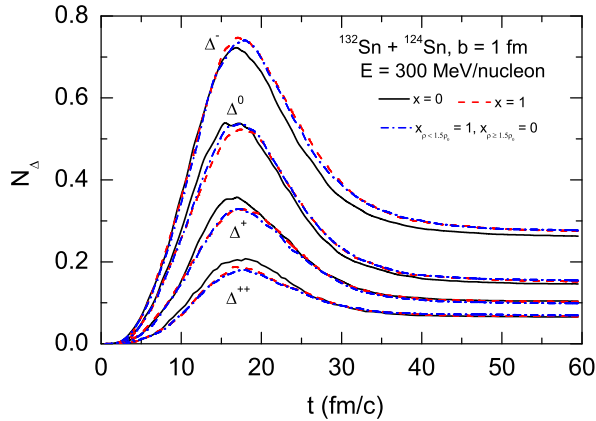}
\end{center}
\caption {(Color online) \label{delta} Production of $\Delta$ resonances as
a function of time with different x parameters in the reaction of $^{132}$Sn + $^{124}$Sn at a
beam energy of 300 MeV/nucleon and an impact parameter of 1 fm.}
\end{figure}
As shown by Eq.~(9), pion production is from the decay of $\Delta$ resonance,
it is thus necessary to also show the production of different $\Delta$ resonances in $^{132}$Sn + $^{124}$Sn
reaction. Fig.~\ref{delta} shows the production of $\Delta$ resonance as
a function of time with different x parameters. It is seen that $\Delta^{-}$ and $\Delta^{++}$ productions are sensitive to the symmetry energy and have opposite effects of the symmetry energy. The symmetry energy above 1.5 times saturation density has neglected effects. This is the reason why $\pi ^{-}/\pi ^{+}$ ratio shown in Fig.~\ref{rpion} is less sensitive to the symmetry energy at densities above 1.5 times saturation density.

\begin{figure}[t]
\begin{center}
\includegraphics[width=0.5\textwidth]{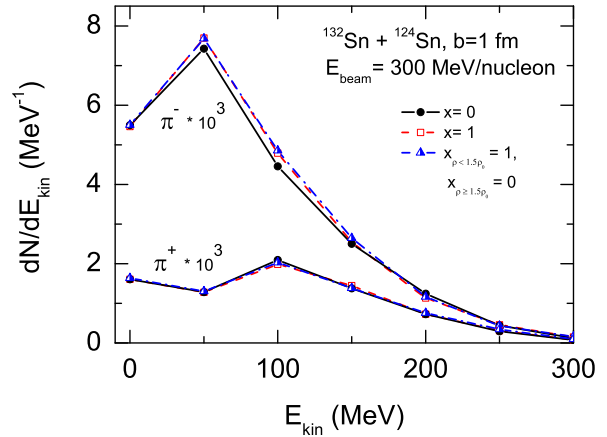}
\end{center}
\caption {(Color online) \label{spect} Kinetic energy spectra of $\pi ^{-}$, $\pi ^{+}$ with different symmetry energies in the reaction of $^{132}$Sn + $^{124}$Sn at a
beam energy of 300 MeV/nucleon and an impact parameter of 1 fm.}
\end{figure}
Fig.~\ref{spect} shows kinetic energy spectra of $\pi ^{-}$, $\pi ^{+}$ with different symmetry energies. It is seen that the kinetic energy spectrum of $\pi ^{-}$ are more sensitive to the symmetry energy than that of $\pi ^{+}$. The kinetic energy spectra of $\pi ^{-}$, $\pi ^{+}$ have opposite effects of the symmetry energy. The soft symmetry energy (x = 1) causes more $\pi ^{-}$
production while it causes somewhat less $\pi ^{+}$ production. As shown in Fig.~\ref{spect}, from the kinetic energy spectrum of $\pi ^{-}$, the effects of the symmetry energy above 1.5 times saturation density can be neglected. Therefore, the $\pi ^{-}/\pi ^{+}$ ratio mainly probes the symmetry energy below 1.5 times saturation nuclear density.

\begin{figure}[t]
\begin{center}
\includegraphics[width=0.5\textwidth]{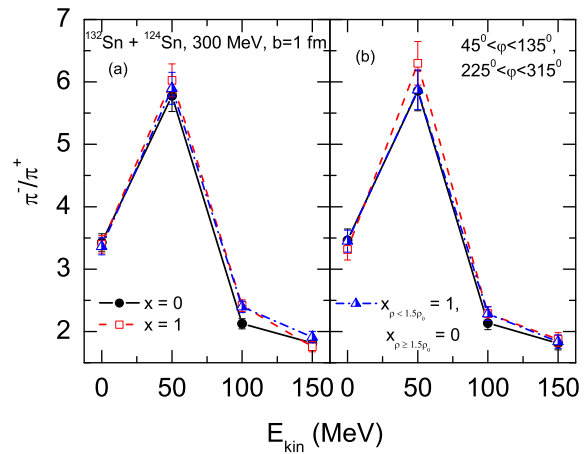}
\end{center}
\caption {(Color online) \label{ekrpion} $\pi ^{-}/\pi ^{+}$ ratio as a function of pion
kinetic energy in central $^{132}\rm {Sn}+^{124}\rm {Sn}$ reaction at 300 MeV/nucleon with (b) and without (a) azimuthal angle cuts.}
\end{figure}
In order to probe the symmetry energy at maximal density reached in heavy-ion collisions,
one way is to reduce the strong interactions in the final state soon after the probe produced
at high density. Besides using electromagnetic probe such as the hard photon \cite{yongp1}, the other way is to make some kinematic cuts such as the yield ratios of squeezed-out nucleons and pions perpendicular to the reaction plane \cite{yongso,gao13}.

Shown in Fig.~\ref{ekrpion} is the $\pi ^{-}/\pi ^{+}$ ratio as a function of pion
kinetic energy in central $^{132}\rm {Sn}+^{124}\rm {Sn}$ reaction at 300 MeV/nucleon with (b) and without (a) azimuthal angle ($45^{0} < \varphi < 135^{0}, 225^{0} < \varphi < 315^{0}$ means the direction perpendicular to the reaction plane) cuts.
From panel (a), one sees that below 150 MeV kinetic energy, the soft symmetry energy (x = 1) corresponds to a high value of the $\pi ^{-}/\pi ^{+}$ ratio while the stiff symmetry energy (x = 0) corresponds to a low value. Without azimuthal angle cuts, the effects of the symmetry energy above 1.5 times saturation density are minor. While with the azimuthal angle cuts shown in panel (b), around 50 MeV, the effects of the symmetry energy above 1.5 times saturation density become larger, even play a major role around the Coulomb peak \cite{ylcz06}. Because the impact parameter in heavy-ion collisions is usually not zero, pions emitted perpendicular to the reaction plane in general have less final strong interactions suffered from low densities. Pions emitted perpendicular to the reaction plane thus more clearly reflect initial pion production at high densities. Therefore, to probe the symmetry energy at maximal density in heavy-ion collisions, it is necessary to make some kinematic cuts for probes that are sensitive to the symmetry energy. From the right panel of Fig.~\ref{ekrpion}, it is seen that within the error bars and the experimental problems in measuring pion observables, it is still difficult
to draw conclusions from comparisons to data. However, one may conclude that the $\pi^{-}/\pi^{+}$ ratio may be able to probe the density-dependent symmetry energy above 1.5 times saturation density by making some kinematic restrictions.

In this regard, most potential observables that are sensitive to the symmetry energy, such as free neutron/proton ratio, $\pi^-/\pi^+$ ratio, t/$^{3}$He ratio, differences of nucleon transverse or elliptic flows, eta production \cite{eta13,xiao14}, etc., are needed to study on what conditions they probe the symmetry energy at maximal density reached in heavy-ion collisions.

\section{Conclusions}

In summary, we simulated the effects of symmetry energy on n/p and $\pi ^{-}/\pi ^{+}$ ratios in $^{132}\rm {Sn}+^{124}\rm {Sn}$ reaction at
300 MeV/nucleon that are being carried out at RIBF-RIKEN in Japan using the
SAMURAI-Time -Project-Chamber. The maximal baryon density reached is about 1.7 times saturation density. The n/p and $\pi ^{-}/\pi ^{+}$ ratios generally probe the symmetry energy below 1.5 times saturation density. However, the $\pi^{-}/\pi^{+}$ ratio may be able to probe the density-dependent symmetry energy above 1.5 times saturation density by making some kinematic restrictions. To achieve the goal of probing the symmetry energy at densities far from saturation density, more clean ways are needed to study on symmetry-energy-sensitive observables.

In Ref.~\cite{xujun2016}, it is shown that even if the main inputs in different transport models are the same, many simulated physical quantities still have evident differences. And since the isospin-sensitive observables are sensitive to the detailed tactics of transport model code, the differences of the isospin-sensitive observables given by different transport models should be even larger. To probe the high-density behavior of nuclear symmetry energy, besides using state-of-the-art inputs in a transport model, some specific details of algorithm design should be also treated seriously.

\section{Acknowledgments}

This work is supported in part by the National Natural
Science Foundation of China under Grants No. 11375239,
No. 11435014, and No. 11275073 and the Fundamental Research
Funds for the Central University of China under Grant
No. 2014ZG0036.


\begin{thebibliography}{100}

\bibitem{li08}B.A. Li, L.W. Chen and C.M. Ko, Phys. Rep. {\bf 464}, 113 (2008).
\bibitem{bar05}V. Baran, M. Colonna, V. Greco, M. Di Toro, Phys. Rep. {\bf 410}, 335 (2005).
\bibitem{pawl2002}P. Danielewicz, R. Lacey, and W.G. Lynch, Science {\bf 298}, 1592 (2002).
\bibitem{Guo14}W.M. Guo, G.C. Yong, Y.J. Wang, Q.F. Li, H.F. Zhang, W. Zuo, Phys. Lett. B {\bf 738}, 397 (2014).

\bibitem{wolter06}C. Fuchs, H.H. Wolter, Eur. Phys. Journal A {\bf 30}, 5 (2006).
\bibitem{xie13}W.J. Xie, J. Su, L. Zhu, F.S. Zhang, Phys. Lett. B {\bf 718}, 1510 (2013).
\bibitem{xiao09}Z.G. Xiao, B.A. Li, L.W. Chen, G.C. Yong, M. Zhang, Phys. Rev. Lett. {\bf 102}, 062502 (2009).
\bibitem{prassa07}V. Prassa, G. Ferini, T. Gaitanos, H.H. Wolter, G.A. Lalazissis, M. Di Toro, Nucl. Phys. A {\bf 789}, 311 (2007).
\bibitem{feng10} Z.Q. Feng, G.M. Jin, Phys. Lett. B {\bf 683}, 140 (2010).
\bibitem{hong2014}J. Hong, P. Danielewicz, Phys. Rev. C {\bf 90}, 024605 (2014).
\bibitem{Reisdorf07}W. Reisdorf, et al., Nucl. Phys. A. {\bf 781}, 459 (2007).
\bibitem{cozma16}M.D. Cozma, arXiv:1603.00664 (2016).

\bibitem{Lat01}J.M. Lattimer, M. Prakash, Ap. J. {\bf 550}, 426 (2001).
\bibitem{Lat04}J.M. Lattimer, M. Prakash, Science {\bf 304}, 536 (2004).
\bibitem{Vil04}Adam R. Villarreal and Tod E. Strohmayer, Ap. J. {\bf 614}, L121 (2004).
\bibitem{Ste05}A.W. Steiner, M. Prakash, J.M. Lattimer, P.J. Ellis, Phys. Rep. {\bf 411}, 325 (2005).
\bibitem{liu15}H.L. Liu, G.C. Yong, and D.H. Wen, Phys. Rev. C {\bf 91}, 044609 (2015).
\bibitem{sep}Symmetry Energy Project, \url{https://groups.nscl.msu.edu/hira/
sepweb/pages/home.html.}
\bibitem{shan15}R. Shane, A.B. McIntosh, T. Isobe, W.G. Lynch, H. Baba, J. Barney, Z. Chajecki,
M. Chartier, J. Estee, M. Famiano, B. Hong, K. Ieki, G. Jhang, R. Lemmon, F. Lua,
T. Murakami, N. Nakatsuka, M. Nishimura, R. Olsen, W. Powell, H. Sakurai,
A. Taketani, S. Tangwancharoen, M.B. Tsang, T. Usukura, R. Wang,
S.J. Yennello, J. Yurkon, Nuclear Instruments and Method A {\bf 784}, 513 (2015).
\bibitem{yong20151}G.C. Yong, arXiv: 1503.08523 (2015).
\bibitem{yong20152}G.C. Yong, Phys. Rev. C {\bf 93}, 044610 (2016).
\bibitem{yong20153}W.M. Guo, G.C. Yong, H. Liu, W. Zuo, Phys. Rev. C {\bf 91}, 054616 (2015).
\bibitem{lyz05}B.A. Li, G.C. Yong, W. Zuo, Phys. Rev. C {\bf 71}, 014608 (2005).
\bibitem{li2003}B.A. Li, Phys. Rev. C {\bf 67}, 017601 (2003).

\bibitem{lim2003}B.A. Li, Champak B. Das, Subal Das Gupta, Charles Gale, Nucl. Phys. A {\bf 735} 563 (2004).

\bibitem{gaoy2011}Y. Gao, L. Zhang, H.F. Zhang, X.M. Chen, G.C. Yong, Phys. Rev. C {\bf 83}, 047602 (2011).
\bibitem{guom2014}W.M. Guo, G.C. Yong, W. Zuo, Phys. Rev. C {\bf 90}, 044605 (2014).
%\bibitem{lib2015}B.A. Li, W.J. Guo, Z. Shi, Phys. Rev. C {\bf 91}, 044601 (2015).
\bibitem{npa2010} W. Reisdorf et al., Nucl. Phys. A {\bf 848}, 366 (2010).

\bibitem{mzhang2009}Ming Zhang, Zhi-Gang Xiao, Bao-An Li, Lie-Wen Chen, Gao-Chan Yong, Sheng-Jiang Zhu, Phys. Rev. C {\bf 80}, 034616 (2009).

\bibitem{xu2000}H.S. Xu, M.B. Tsang, T.X. Liu, X. D. Liu, W.G. Lynch, W.P. Tan, A. Vander Molen, G. Verde, A. Wagner, H.F. Xi, C.K. Gelbke, L. Beaulieu, B. Davin, Y. Larochelle, T. Lefort, R. T. de Souza, R. Yanez, V. E. Viola, R. J. Charity and L. G. Sobotka,
    Phys. Rev. Lett. 85, 716 (2000).
\bibitem{stock}R. Stock, Phys. Rep. {\bf 135}, 259 (1986).
\bibitem{ylcz06}G.C. Yong, B.A. Li, L.W. Chen, W. Zuo, Phys. Rev. C {\bf 73}, 034603 (2006).
\bibitem{ono16}Natsumi Ikeno, Akira Ono, Yasushi Nara, Akira Ohnishi, Phys. Rev. C {\bf 93}, 044612 (2016).

\bibitem{yongp1}G.C. Yong, B.A. Li, and L.W. Chen, Phys. Lett. B {\bf 661}, 82 (2008).
\bibitem{yongso}G.C. Yong, B.A. Li, and L.W. Chen, Phys. Lett. B {\bf 650}, 344 (2007).
\bibitem{gao13}Y. Gao, G.C. Yong, Y.J. Wang, Q.F. Li, W. Zuo, Phys. Rev. C {\bf 88}, 057601 (2013).
\bibitem{eta13}G.C. Yong, B.A. Li, Phys. Lett. B {\bf 723}, 388(2013).

\bibitem{xiao14}Z.G. Xiao, G.C. Yong, L.W. Chen, B.A. Li, M. Zhang, G.Q. Xiao, N. Xu,
Eur. Phys. Journal A {\bf 50}, 37 (2014).
\bibitem{xujun2016}Jun Xu, Lie-Wen Chen, ManYee Betty Tsang, Hermann Wolter, Ying-Xun Zhang, Joerg Aichelin, Maria Colonna, Dan Cozma, Pawel Danielewicz, Zhao-Qing Feng, Arnaud Le Fevre, Theodoros Gaitanos, Christoph Hartnack, Kyungil Kim, Youngman Kim, Che-Ming Ko, Bao-An Li, Qing-Feng Li, Zhu-Xia Li, Paolo Napolitani, Akira Ono, Massimo Papa, Taesoo Song, Jun Su, Jun-Long Tian, Ning Wang, Yong-Jia Wang, Janus Weil, Wen-Jie Xie, Feng-Shou Zhang, Guo-Qiang Zhang, Phys. Rev. C {\bf 93}, 044609 (2016).


\end{thebibliography}
\end{document}